\documentclass[10pt]{article}

\usepackage{accents,verbatim}
\usepackage{latexsym}
\usepackage{graphicx}
\usepackage{amssymb,wasysym}
\usepackage{natbib,color,mathrsfs}

\usepackage[labelfont=bf,labelsep=period,justification=raggedright]{caption}

\date{}

\newcommand{\E}{\mathbb{E}}

\newcommand{\radius}{\rho}
\newcommand{\error}{\varepsilon}
\newcommand{\xmass}{\bfx^\bullet}
\newcommand{\vmass}{v^\bullet}

\newcommand{\hess}{\mathbf{H}}
\newcommand{\vobs}{v_{\mathrm{obs}}}
\newcommand{\wobs}{w_{\mathrm{obs}}}
\newcommand{\kobs}{k_{\mathrm{obs}}}
\newcommand{\bfxpert}{\mathbf{x}_{\mathrm{perturbed}}}
\newcommand{\bfxnf}{\mathbf{x}_{\mathrm{noise\mbox{-}free}}}

\newcommand{\bfb}{\mathbf{b}}

\newcommand{\bff}{\mathbf{f}}

\newcommand{\bfx}{\mathbf{x}}

\newcommand{\bfA}{\mathbf{A}}

\newcommand{\bfF}{\mathbf{F}}

\newcommand{\bfK}{\mathbf{K}}

\newcommand{\expect}{\mathbb{E}}

\newcommand{\gV}{g_V}
\newcommand{\EV}{E_V}
\newcommand{\gNa}{G_{Na}}
\newcommand{\ENa}{E_{Na}}
\newcommand{\gCa}{g_{Ca}}
\newcommand{\ECa}{E_{Ca}}
\newcommand{\Vconst}{V_{const}}

\begin{document}

\begin{flushleft}
{\Large
\textbf{How synchronization protects from noise}
}
\\
Nicolas Tabareau$^{1,\ast}$, 
Jean-Jacques Slotine$^{2}$, 
Quang-Cuong Pham$^{1}$
\\
\bf{1} LPPA, Coll\`ege de France, Paris, France
\\
\bf{2} Nonlinear Systems Laboratory, MIT, Cambridge, MA 02139, USA
\\
$\ast$ E-mail: nicolas.tabareau@gmail.com
\end{flushleft}

\section*{Abstract}

Synchronization phenomena are pervasive in biology. In neuronal
networks, the mechanisms of synchronization have been extensively
studied from both physiological and computational viewpoints. The
functional role of synchronization has also attracted much interest
and debate. In particular, synchronization may allow distant sites in
the brain to communicate and cooperate with each other, and therefore
it may play a role in temporal binding and in attention and
sensory-motor integration mechanisms. 
In this article, we study another role for synchronization: the
so-called "collective enhancement of precision." We argue, in a full
nonlinear dynamical context, that synchronization may help protect
interconnected neurons from the influence of random perturbations --
intrinsic neuronal noise -- which affect all neurons in the nervous
system.  This property may allow reliable computations to be carried
out even in the presence of significant noise (as experimentally found
e.g., in retinal ganglion cells in primates), as mathematically it is
key to obtaining meaningful downstream signals, whether in terms of
precisely-timed interaction (temporal coding), population coding, or
frequency coding. Using stochastic contraction theory, we show how
synchronization of nonlinear dynamical systems helps protect these
systems from random perturbations.
Our main contribution is a mathematical proof that, under specific
quantified conditions, the impact of noise on each individual system
and on the spatial mean can essentially be cancelled through
synchronization. Similar concepts may be applicable to questions in
systems biology.

\section*{Author Summary}
Synchronization phenomena are pervasive in biology, creating
collective behavior out of local interactions between neurons, cells,
or animals. Many of these interactions occur in the presence of large
amounts of noise or disturbances, making one wonder how meaningful
behavior can arise in these highly uncertain conditions. In this paper
we show mathematically, in a very general context, that
synchronization is not just robust to the deleterious effects of noise
and disturbances: it actually {\it protects} subsystems from them. In
effect, synchronization makes subsystems work as a team in fighting
noise and successfully achieving their target behavior.  One mechanism
for strong synchronization is for subsystems to jointly create and
then share a common signal, such as a mean electrical field or a mean
chemical concentration, in effect making each subsystem directly
connected to all others. Conversely, extracting meaningful information
from average measurements over populations of cells (as commonly used
for instance in electro-encephalography, or more recently in
brain-machine interfaces) may require the presence of synchronization
mechanisms similar to those we describe.

\section*{Introduction}

Synchronization phenomena are pervasive in biology. In neuronal
networks \cite{Sin93,Buz06,TieX08}, a large number of studies have
sought to unveil the mechanisms of synchronization, from both
physiological \citep{HG05,FukX06} and computational
viewpoints\citep[see for instance][and references therein]{PS07}. In
addition, the {\it functional} role of synchronization has also
attracted considerable interest and debates. In particular,
synchronization may allow distant sites in the brain to communicate
and cooperate with each other \citep{CK05,CanX06,WomX07} and therefore
may play a role in temporal binding \citep{Gro00,ES01} and in attention
and sensory-motor integration mechanisms
\citep{WF07,PalX07,gregoriou09}.

In this article, we study another role for synchronization: the
so-called \emph{collective enhancement of precision} \citep[see
e.g.][]{sherman1988,sherman1991,kinard1999mbp}, an intuitive and often
quoted phenomenon with comparatively little formal
analysis~\citep{winfree2001gbt}.  We explain mathematically why
synchronization may help {\it protect} interconnected nonlinear
dynamic systems from the influence of random perturbations. In the
case of neurons, these perturbations would correspond to so-called
``intrinsic neuronal noise'' \citep[see for instance][]{FaiX08}, which
affect all of the neurons in the nervous system. In the presence of
significant noise intensities (as experimentally found in
e.g. retinal ganglion cells in primates \citep{CroX93}), this property
would be required for meaningful and reliable computations to be
carried out.


In general, the behavior of a nonlinear dynamical system can be
dramatically affected by the presence of noise, as e.g. in chaotic
systems. However it has been shown analytically that some limit-cycle
oscillators commonly used as simplified neuron models, such as
FitzHugh-Nagumo (FN) oscillators, are basically unperturbed when they
are subject to a small amount of white noise \citep{TR98}. Yet, a
larger amount of noise breaks this ``robustness'', both in the state
space and in the frequency space (Figures~1(A)-(D)). This suggests that
both temporal coding and frequency coding may be unusable in the
context of large neuronal noise.

One might argue that it could be possible to recover some information
from the noisy FN oscillators by considering the activities of a large
number of oscillators
\emph{simultaneously}~\citep{DA01,FaiX08}. Figure~2(A) shows
that the spatial mean of the noisy oscillators still carries very
little information when the noise intensities are large, making the
population coding hypothesis also unlikely in this context. In other words,
if the underlying dynamics are fundamentally {\it nonlinear},
as in the case of our FN oscillators, the spatial mean of the signals is 
``clean,'' but the nonlinear nature of the systems dynamics prevents
the familiar ``averaging out'' of noise through multiple measurements.

By contrast, one can observe that when oscillators are {\it
synchronized} through mutual couplings, then they become ``protected''
from noise, whether in temporal (Figure~1(E)), frequential (Figure~1(F))
or ``populational'' aspects (Figure~2(B)). Thus, in some sense, the
linear effect of averaging noise while preserving signal \citep{Gel74}
can be achieved for these highly nonlinear dynamic components {\it
through the process of synchronization}. Our aim in this article is to
give mathematical elements of explanation for this phenomenon, in a
full nonlinear setting. It is also to 
suggest elements of response to a more general question, namely: what
is the precise {\it meaning} of ensemble measurements or population
codes, and what information do they convey about the underlying
dynamics and signals?

\section*{Results}

\subsection*{General analytical result}

Consider a diffusive network of $d$-dimensional noisy non-linear
dynamical systems
\begin{equation}
  \label{eq:main}
  d\bfx_i=\left(\bff(\bfx_i,t)+\sum_{j\neq
    i}\bfK_{ji}(\bfx_j-\bfx_i)\right)dt + \sigma dW_i,\ i=1\dots n
\end{equation}
where $\bff=(f_1,\dots,f_d)^T$ is a $\mathbb{R}^d\to\mathbb{R}^d$
function. Note that the noise intensity $\sigma$ is intrinsic to the
dynamical system (i.e. independent of the inputs), which is consistent
with experimental findings \citep{CroX93}. For simplicity, we set
$\sigma$ to be a constant in this article, although the case of time-
and state-dependent noise intensities can be easily adapted
from~\citep{PhaX07}.

We consider four hypotheses that will enable us to relate the
trajectory of any noisy element of the network $\bfx_i$ to the
trajectory of the noise-free system $\bfxnf$ driven by equation
\[
d\bfxnf=\bff(\bfxnf,t)dt.
\]
Hypothesis (H1)
is an assumption on the form of the network. (H2) gives a bound on the
nonlinearity of the dynamics $\bff$. (H3) states that $\bff$ is robust
to small perturbations. Finally, (H4) requires that the dynamical
systems in the network are synchronized.

\begin{description}
\item[(H1)] The network is balanced, that is, for any element of the
  network, the sum of the incoming connection weights equals the sum
  of the outgoing connection weights
  \[
  \forall i\quad \sum_{j}\bfK_{ji}=\sum_{j}\bfK_{ij}.
  \]
  In particular, any symmetric network is balanced.

\item[(H2)] Let $\hess_j$ denote the Hessian matrix of the function
  $f_j$ and let $\lambda_{\max}(\hess_j)$ denote its largest
  eigenvalue. For all $j$, we assume that $\lambda_{\max}(\hess_j)$ is
  uniformly upper-bounded by a constant
  $\frac{1}{\sqrt{d}}\|\hess\|$. This implies in particular that
  \[
  \forall \bfx,j,t \quad \bfx^T \hess_j \bfx \leq
  \frac{\|\hess\|}{\sqrt{d}} \| \bfx \|^2.
  \]
  This hypothesis gives us a bound on the nonlinearity of $\bff$, the
  extreme case being $\|\hess\| = 0$ for a linear system.

\item[(H3)] The dynamics $\bff$ is robust to small perturbations. More
  precisely, consider two systems starting from the same initial
  conditions but driven by slightly different dynamics
  \[
  \dot{\bfx}_\mathrm{noise-free}=\bff(\bfxnf,t)
  \]
  and
  \[
  \dot{\bfx}_\mathrm{perturbed}=\bff(\bfxpert,t)+P,
  \]
  then $\ \expect(\|P\|)\to 0\ $ implies $\ \|\bfxnf-\bfxpert\|\to 0$.

  In particular, such a property has been demonstrated in the case of
  FN oscillators, with $P$ representing a white noise process
  \citep{TR98}.

\item[(H4)] After exponential transients, the expected sum of the
  squared distances between the states of the elements of the
  network is bounded by a constant $\rho$
  \[
  \expect\left(\sum_{i<j}\|\bfx_i-\bfx_j\|^2\right)\leq \rho.
  \]
  This is where synchronization will come into play, because
  synchronization is an effective way to reduce the bound~$\rho$. Some
  precise conditions for this will be given later.
\end{description}
We show in {\bf Methods} that under these hypotheses and when
$n\to\infty$ and $\rho / n^2 \to 0$, the distance between the
trajectory of any noisy element $\bfx_i$ of the network and that of
the noise-free system $\bfxnf$ tends to zero,
with the impact of noise on the mean trajectory evolving as
\[
\frac{\radius\|\hess\|}{2n^2}\ + \ \frac{\sigma}{\sqrt{n}} .
\]
In particular, when $\bff$ is a time-varying linear system of the form
$\bff(\bfx,t) = \bfA(t) \bfx + \bfb(t)$, we recover the known
result~\citep{enright1980} that the impact of noise evolves as the
inverse square root of $n$.
More generally, linear components of the system dynamics (including,
in particular, the input signals) do not contribute to the first term
of the above upper bound.

\subsection*{Synchronization in networks of noisy FN oscillators}

We now give conditions to guarantee hypothesis (H4) for all-to-all
networks of FN oscillators with identical couplings. The dynamics of
$n$ noisy FN oscillators coupled by (gap-junction-like) diffusive
connections is given by
\begin{equation} \label{eq:FN general}
\left\{
\begin{array}{rcl}
  dv_i & = & \left(c f(v_i,w_i,I)+\sum_j \frac{k}{n} (v_j-v_i)\right)dt 
  + \sigma dW_i \\
  dw_i & = & -\frac{1}{c}(v_i-a +b w_i)dt
  \end{array}
\right.
\end{equation}
where $f(v,w,I)=v-\frac{1}{3}v^3+w+I$.
%
%
%
We show in {\bf Methods} that, after transients of rate $k$,
\begin{equation}
  \label{eq:bound-diff}
  \expect\left(\sum_{i<j}(v_i-v_j)^2\right)\leq \frac{n(n-1)\sigma^2}{k}.  
\end{equation}
Thus, (H4) is verified with 
\begin{equation}
  \label{eq:rho}
  \rho=\frac{n(n-1)\sigma^2}{k}. 
\end{equation}
For large $n$, we have $\rho/n^2\sim \sigma^2/k$, which converges to 0
when $k\to\infty$ (see Figure~3(A)).

Hypothesis (H1) is also verified because an all-to-all network with
identical couplings is symmetric, therefore balanced. 
Since the $(v_i,w_i)^T$ are oscillators with stable limit cycles, it
can be shown that the trajectories of the $v_i$ are bounded by a
common constant $M$. Thus (H2) is verified with
$\|\hess\|=2cM$. Finally, (H3) may be adapted from
\citep{TR98}. Indeed, we believe that the arguments of \citep{TR98}
can be extended to the case of non-white noise. Making this point
precise is the subject of ongoing work.

Using now the ``general analytical result'', we obtain that, given any (non
necessarily small) noise intensity $\sigma$, in the limits for
$k\to\infty$ and $n\to\infty$ and after exponential transients, the
behavior of any oscillator will be arbitrary close to that of a
noise-free oscillator (Figure~1).

This statement can be further tested by constructing a model-based
nonlinear state estimator (observer) \citep{LS98}. Let $(v_i,w_i)^T$
be a noisy synchronized oscillator and consider the observer
\begin{equation}
  \label{eq:obs}
  \left\{
    \begin{array}{rcl}
      \vobs & = & c f(\vobs,\wobs,I)+\kobs(v_i-\vobs) \\
      \wobs & = & -\frac{1}{c}(\vobs-a +b \wobs).
    \end{array}
  \right.
\end{equation}

If $v_i$ has the same trajectory as a noise-free FN oscillator, then
it can be shown that $(\vobs,\wobs)^T$ tends exponentially to
$(v_i,w_i)^T$, independently of the observer's initial conditions
\citep{LS98}. Thus the squared distance $(\vobs-v_i)^2$ indicates how
close $v_i$ is from a noise-free oscillator (Figure~3(B)).

\subsection*{Some extensions and other examples}

We provide in this section some ideas and simulation results which
show the genericity of the concepts presented above. However, the full
mathematical explanation for some results is still under development.

\paragraph{Probabilistic networks.}  
In practice, all-to-all neuronal networks of large size are
rare. Rather, the mechanisms of neuronal connections in the brain are
believed to be probabilisitic in nature \citep[see][for a
review]{Stro01}. Here, we consider a probabilistic symmetric network
of $n$ oscillators, where any pair of oscillators has probability $p$
to be symmetrically connected and probability $1-p$ to be
unconnected. Figure~4 shows simulation results for $p=0.1$.

\paragraph{Quorum sensing.}

In system biology, quorum sensing \citep{GarX04,taylor2009dqs}), where
individual cells measure mean chemical concentrations in their
environment, may be seen as a mechanism implementing all-to-all
coupling. For instance, assuming that the mean value of the $\bfx_i$'s
can be provided by the environment as $ \xmass= \frac{1}{n} \sum_i
\bfx_i, $ then, the all-to-all network~(\ref{eq:main}) with $\bfK_{ji}
= \mathscr{K}_i $ can be written as a star network where damping is
added locally and each cell $\bfx_i$ is only connected to the common
signal
\begin{equation} \label{eq:quorum sensing}
  d\bfx_i = \left(\bff(\bfx_i,t) + n {\mathscr{K}}_i (\xmass - \bfx_i)\right)dt 
  + \sigma dW_i.
\end{equation}
Quorum sensing, and more generally the measurement of a common mean
signal, can thus be seen as a practical (and biological plausible) way
to implement all-to-all coupling with $2 n$ connections instead of
$n^2$.

\paragraph{Hindmarsh-Rose oscillators.} Hindmarsh-Rose oscillators are
three-dimensional dynamical systems that are also often used as neuron
models
\[
\left\{
\begin{array}{lcl}
  dV & = & (I - n - m - V^3 + \gV V + \EV V^2) dt + \sigma dW \\
  dn & = & (\gNa+\ENa V^2-n) dt \\
  dm & = & (\gCa (\ECa (V+\Vconst)-m)) dt
  \end{array}
\right.
\]
with $\gV = 0.5$; $\EV = 2.8$; $\gNa = 0$; $\ENa = 4.4$; $\gCa =
0.001$; $\ECa = 9$; $\Vconst = 7/ECa$.
%
%
These oscillators can exhibit more complex behaviors (including
spiking and bursting regimes \citep{Izh04}) than FitzHugh-Nagumo
oscillators. The proofs of (H3) and (H4) for Hindmarsh-Rose
oscillators are the object of ongoing research.

We made the inputs time-varying in this simulation. In fact, all the
previous calculations can be straightforwardly extended to the case of
time-varying inputs, as long as those inputs are the same for all the
oscillators \citep[see][]{PS07}.

One can observe from the simulations (see Figure~5) that the
synchronized oscillators let the inputs' signal through, while the
uncoupled oscillators completely blur the signal out.

\section*{Discussion}

We have argued that synchronization may represent a fundamental
mechanism to protect neuronal assemblies from noise, and have
quantified this hypothesis using a simple nonlinear neuron model.
This may further strengthen our understanding of synchronization in
the brain as playing a key functional role, rather than as being
mostly an epiphenomenon.

It should be noted that the causal relationship studied here -- effect
of synchronization on noise -- is converse to one usually investigated
formally in the literature -- effect of noise on synchronization:
destructive effect \citep{TK01}; constructive effect
\citep{MS95,TT04}; for a review, see \citep{ErmX08}. Also, previous
papers have studied a similar phenomenon of improvement in precision
by synchronization.
Enright~\citep{enright1980} shows $\sqrt{N}$ improvement in a model of
coupled relaxation oscillators, all interacting through a common
accumulator variable (possibly being the pineal gland).
This $\sqrt{n}$ improvement has been experimentally shown in real
heart cells~\citep{clay1979}. More recently, \citep{NTS01} shows a way
to get better than $\sqrt{n}$ improvement.
However, their studies primarily focused on the case
of phase oscillators, which are linear dynamical systems. In contrast,
we concentrate here on the more general case of nonlinear oscillators,
and quantify in particular the effect of the oscillators'
nonlinearities. The hypotheses we consider are also different: while
most existing approaches (including \citet{NTS01}) assume weak
couplings and small noise intensities, we consider here strong
couplings and arbitrary noise intensities.

The mechanisms highlighted in the present paper may also underly other
types of ``redundant'' calculations. In otoliths for instance, ten of
thousands of hair cells jointly compute the three components of
acceleration~\citep{KanX00,EA04}. In muscles, thousands of individual
fibers participate in the control of one single degree of
freedom. Similar questions may also arise in systems biology, e.g., in
cell mechanisms of quorum sensing where individual cells measure mean
chemical concentrations in their environment in a fashion functionally
similar to all-to-all coupling~\citep{GarX04,taylor2009dqs}, in
mechanical coupling of motor proteins~\citep{hendricks2009cdk}, in the
context of transcription-regulation networks~\citep{Alo07}, and in
differentiation dynamics~\citep{suel2007tan}.

Finally, the results point to the general question: what is the
precise meaning of ensemble measurements or population codes, what
information do they convey about the underlying signals, and is the
presence of synchronization mechanisms (gap-junction mediated or
other) implicit in this interpretation? As such, they may also shed
light on a somewhat ``dual'' and highly controversial current issue. Ensemble
measurements from the brain can correlate to behavior, and they have
been suggested e.g. as inputs to brain-machine interfaces.  Are these
ensemble signals actually available to the brain~\citep{fregnac},
perhaps through some process akin to quorum sensing, and therefore
functionally similar to (local) all-to-all coupling? Are local field
potentials~\citep{pesaran2002tsn} plausible candidates for a role in this
picture?

\section*{Methods}

\subsection*{Proof of the general analytical result}

In the noise-free case ($\sigma=0$), it can be shown that, for strong
enough coupling strengths, the elements of the network synchronize
completely, that is, after exponential transients, we have $\rho=0$ in
(H4) \citep{PS07}. Thus, all the $\bfx_i$ tend to a common trajectory,
which is in fact a nominal trajectory of the noise-free system
$\dot{\bfx}_{\mathrm{noise-free}}=\bff(\bfxnf,t)$, because all the
couplings vanish on the synchronization subspace.

In the presence of noise, it is not clear how to relate the trajectory of
each $\bfx_i$ to a nominal trajectory of the noise-free
system. Nevertheless, we still know that the $\bfx_i$ live ``in a
small neighborhood'' of each other, as quantified by (H4). Thus, if
the center of this small neighborhood follows a trajectory similar to
a nominal trajectory of the noise-free system, then one may gain some
information on the trajectories of the~$\bfx_i$.

To be more precise, let $\xmass$ be the center of mass of the
$\bfx_i$, that is
\begin{equation}
  \xmass= \frac{1}{n} \sum_i \bfx_i.
\end{equation}
Observe that, after expansion and rearrangement, the sum
$\sum_{i<j}\|\bfx_i-\bfx_j\|^2$ can be rewritten in terms of the
distances of the $\bfx_i$ from $\xmass$
\[
\sum_{i<j}\|\bfx_i-\bfx_j\|^2
= n \sum_i\|\bfx_i-\xmass\|^2.
\]
Using (H4) then leads to 
\begin{equation} \label{eq:bound general}
  \expect\left(\sum_i\|\bfx_i-\xmass\|^2\right) \leq \frac{\rho}{n}.
\end{equation}
Summing over $i$ the equations followed by the $\bfx_i$ and using
hypothesis (H1), we have
\begin{equation} \label{eq:xmass-dynamic}
d\xmass = \frac{1}{n}\left(\sum_{i} \bff(\bfx_i,t)\right) dt +
\frac{1}{n}\sum_i \sigma dW_i.
\end{equation}
We now make the dynamics explicit with respect to $\xmass$ by letting
\begin{equation}
  \error=\frac{1}{n}\left(\sum_{i=1}^n\bff(\bfx_i,t)\right)-\bff(\xmass,t)
\end{equation}
so that equation~(\ref{eq:xmass-dynamic}) can be rewritten as
\begin{equation} \label{eq:xmass}
  d\xmass = \left(\bff(\xmass,t)-\error \right) dt +
  \frac{1}{n}\sum_i \sigma dW_i.
\end{equation}
Using the Taylor formula with integral remainder, we have
\begin{equation} \nonumber
  \label{eq:xi}
  \begin{array}{cl}
    & f_j(\bfx_i,t) - f_j(\xmass,t) - \bfF_j(\xmass,t)^T
    (\bfx_i-\xmass) \\
    = & \int_0^1 (1-s)(\bfx_i-\xmass)^T \hess_j((1-s)\bfx_i + s \xmass)
    (\bfx_i-\xmass) ds
  \end{array}
\end{equation}
where $\bfF_j$ is the gradient of $f_j$ or, equivalently, the
$j^{\mathrm{th}}$ vector of the Jacobian matrix of $\bff$.
Summing Equation (\ref{eq:xi}) over $i$ and using hypothesis (H2), we
get
\begin{equation} \nonumber
   \label{eq:j} 
  | \sum_i (f_j(\bfx_i,t) - f_j(\xmass,t))| \leq 
  \frac{ \|\hess\|}{2\sqrt{d}} \sum_i \|\bfx_i-\xmass\|^2.
\end{equation}
Summing now inequality (\ref{eq:j}) over $j$ and using inequality
(\ref{eq:bound general}), we get
\begin{equation}
  \label{eq:epsilon} 
  \expect(\|\error\|) 
  \leq\frac{\radius\|\hess\|}{2n^2}
\end{equation}
which implies that $\E(\|\error\|)\rightarrow 0$ when ${\rho}/{n^2}
\rightarrow 0$.

Turning now to the noise term $\frac{1}{n}\sum_i \sigma dW_i$ in
Equation (\ref{eq:xmass}), we have
\begin{equation}
  \label{eq:lgn}
  \frac{1}{n}\sum_i \sigma dW_i \cong \frac{\sigma}{\sqrt{n}} dW
\end{equation}
since the intrinsic noises of the elements of the network are mutually
independent.

Thus, for a given (even large) noise intensity $\sigma$, the
difference between the \emph{dynamics} followed by $\xmass$ and the
noise-free dynamics $\bff$ tends to zero when $n\to\infty$ and $\rho /
n^2 \to 0$. Hypothesis (H3) then implies that $\|\xmass-\bfxnf\|\to
0$. Furthermore, the impact of noise on the mean trajectory evolves as
\begin{equation}
\frac{\radius\|\hess\|}{2n^2}\ + \ \frac{\sigma}{\sqrt{n}} .
\end{equation}
Finally, Equation (\ref{eq:bound general}) and the triangle
inequality
\begin{equation}
  \label{eq:triangle}
  \|\bfxnf-\bfx_i\|\leq
  \|\bfxnf-\xmass\| + \|\xmass-\bfx_i\|
\end{equation}
imply that the \emph{trajectory} of any synchronized element of the
network $\bfx_i$ and that of the noise-free system $\bfxnf$ are also
similar (compare Figure~1(A) and Figure~1(E)).

\subsection*{FN oscillators in an all-to-all network}

\paragraph{Two FN oscillators.}

Consider first the case of two coupled FN oscillators driven by
Equation (\ref{eq:FN general}). Construct the following auxiliary
system (or virtual system, in the sense of ~\citep{WS05}), where $v_1$
and $v_2$ are considered as \emph{external inputs}
\begin{equation}
  \label{eq:auxiliary}
  \left\{\begin{array}{rcl}
      dx_1&=&\left((c-(v_1^2+v_1v_2+v_2^2)-k)x_1+ cx_2)\right)dt \\
      &+&\sqrt{2}\sigma dW \\
      dx_2&=&\left(-\frac{1}{c} x_1 - \frac{b}{c} x_2\right)dt.
    \end{array}\right.
\end{equation}
Remark that $(x_1,x_2)^T=(v_1-v_2,w_1-w_2)^T$ is a particular trajectory
of this system.

Let $\lambda_1=k+(v_1^2+v_1v_2+v_2^2)-c$ and $\lambda_2=b/c$. Assume
that the coupling strength is significantly larger than the system's
parameters, i.e.  $k\gg c$, $k\gg 1/c$ and $k\gg b/c$. Since
$v_1^2+v_1v_2+v_2^2$ is nonnegative for any $v_1$ and $v_2$, we have
either $\lambda_1\geq k$ or $\lambda_1\simeq k$, depending on the
actual value of $v_1^2+v_1v_2+v_2^2$. This implies in particular that
$\lambda_1\gg c$, $\lambda_1\gg 1/c$ and $\lambda_1\gg \lambda_2=b/c$.

Given these asymptotes, the evolution matrix of system
(\ref{eq:auxiliary}) is diagonalizable with eigenvalues $-\lambda_1'$
and $-\lambda_2'$ and eigenvectors respectively $(\lambda_1'',1/c)^T$
and $(c,\lambda_1''')^T$, where
$\lambda_i\simeq\lambda_i'\simeq\lambda_i''\simeq\lambda_i'''$
$(i=1,2)$.

Define now 
\begin{equation}
  \label{eq:x->y}
\left\{\begin{array}{l}
    y_1=\lambda_1'' x_1 + \frac{1}{c} x_2\\
    y_2=c x_1 + \lambda_1'''x_2
\end{array}\right.
\end{equation}
leading to
\[
\left\{\begin{array}{l}
    dy_1=-\lambda_1' y_1dt + \sqrt{2} \sigma \lambda_1'' dW \\
    dy_2=-\lambda_2' y_2dt + \sqrt{2} \sigma c dW.
\end{array}\right.
\]

Since these equations are in fact uncoupled, they can be solved
independently. Using the stochastic contraction results (corollary~1
of \citep{PhaX07}) and the approximations
$\lambda_i\simeq\lambda_i'\simeq\lambda_i''$, this yields
\[
\left\{\begin{array}{l}
    \expect(y_1^2)\leq \sigma^2\lambda_1, \quad
    \textrm{after transients of rate } \lambda_1 \\
    \expect(y_2^2)\leq \frac{c^2\sigma^2}{\lambda_2}, \quad
    \textrm{after transients of rate } \lambda_2.
\end{array}\right.
\]

These bounds can be translated back in terms of the $x_i$ by
inverting~(\ref{eq:x->y})
\[
\left\{\begin{array}{l}
    x_1\simeq \frac{1}{\lambda_1} y_1 -
    \frac{c}{\lambda_1^2}y_2 \\
    x_2\simeq - \frac{1}{c\lambda_1^2}y_1 + \frac{1}{\lambda_1}
    y_2. 
\end{array}\right.
\]

Thus, after transients of rate $\lambda_1$,
\[
  \expect(x_1^2)\leq \frac{\sigma^2}{\lambda_1} \qquad  \ \ \ \ \ 
  \expect(x_2^2)\leq \frac{\sigma^2c^2}{\lambda_1^2\lambda_2}.
\]

Since $(v_1-v_2,w_1-w_2)^T$ is a particular trajectory of system
(\ref{eq:auxiliary}) as we remarked earlier, one finally obtains that,
after transients of rate $k$,
\begin{equation} \label{eq:bound simple case}
\expect((v_1-v_2)^2)\leq \frac{\sigma^2}{k} \qquad \ \ 
\expect((w_1-w_2)^2)\leq \frac{\sigma^2c^3}{bk^2}.
\end{equation}

\paragraph{General case.}

Consider now an all-to-all network with identical couplings as in
Equation (\ref{eq:FN general}). Construct as above the following
$n(n-1)$ auxiliary systems indexed by $(i,j)\in[1\dots n]^2$, where
the $v_i$ are considered as external inputs
\[  \left\{
    \begin{array}{rcl}
      dv_{ij} & = & \left((c-(v_i^2+v_iv_j+v_j^2)-k)v_{ij}+
        cw_{ij})\right)dt \\
      && + \sqrt{2}\sigma dW \\
      dw_{ij} & = & \left(-\frac{1}{c} v_{ij} - \frac{b}{c} w_{ij}\right)dt.
    \end{array}
  \right.
\]
%
Remark that, similarly to the case of two oscillators,
$\left((v_{ij},w_{ij})^T\right)_{i,j}=\left((v_i-v_j,w_i-w_j)^T\right)_{i,j}$
is a particular solution of these equations.
Remark also that each pair $(v_{ij},w_{ij})$ is in fact uncoupled with
respect to other pairs. This allows us to use (\ref{eq:bound simple
  case}) to obtain that, after transients of rate $k$,
\[
\forall i,j,\ i\neq j, \quad \expect((v_i-v_j)^2)\leq \frac{\sigma^2}{k}.
\]
Summing over the $i,j$ yields
\begin{equation}
  \label{eq:bound-diff}
  \expect\left(\sum_{i<j}(v_i-v_j)^2\right)\leq \frac{n(n-1)\sigma^2}{k}.  
\end{equation}
Thus, (H4) is verified with 
\begin{equation}
  \label{eq:rho}
  \rho=\frac{n(n-1)\sigma^2}{k}. 
\end{equation}
For large $n$, we have $\rho/n^2\sim \sigma^2/k$, which converges to 0
when $k\to\infty$ (see Figure 3(A)).

Hypothesis (H1) is also verified because an all-to-all network with
identical couplings is symmetric, therefore balanced. As for (H2),
observe that $\hess_w=0$ and
\[
\hess_v=
\left(\begin{array}{cc}
  2cv & 0\\
  0 &0
\end{array}\right).
\]
Since the $(v_i,w_i)^T$ are oscillators with stable limit cycles, it
can be shown that the trajectories of the $v_i$ are bounded by a
common constant $M$. Thus (H2) is verified with
$\|\hess\|=2cM$. Finally, (H3) may be adapted from
\citep{TR98}. Indeed, we believe indeed that the arguments of
\citep{TR98} can be extended to the case of non-white noise. Making
this point precise is the subject of ongoing work.
%




\section*{Acknowledgments}

JJS is grateful to Uri Alon for stimulating discussions on possible
relevance of the results to cell biology.

%



\bibliography{noise-sync}

\section*{Figure Legends}

\begin{figure}[h!] 
  \centering
  \includegraphics[width=0.99\linewidth]{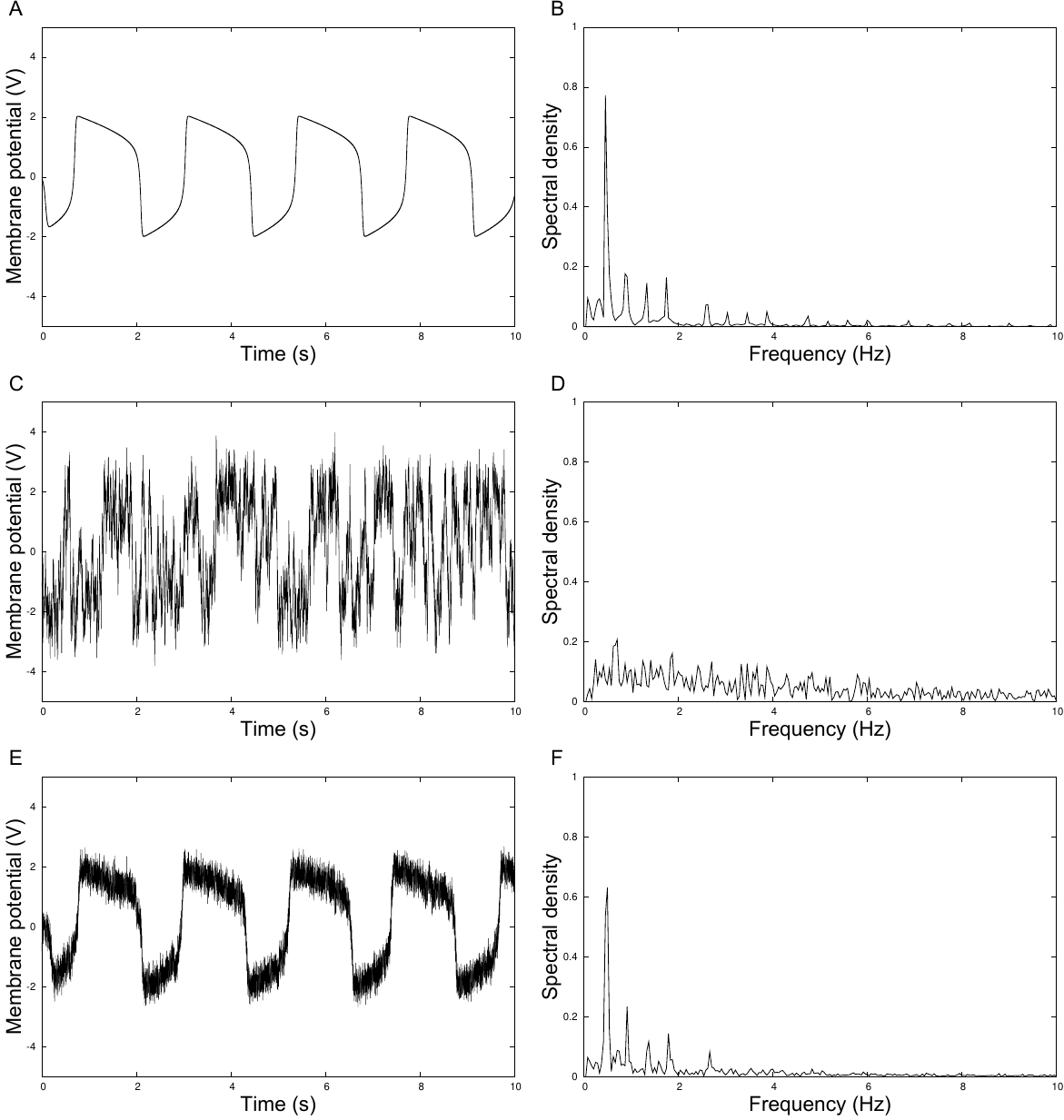}
  \caption{%
    {\bf Simulations of a network of FN oscillators using the
      Euler-Maruyama algorithm \citep{Hig01}.}  
    The dynamics of coupled FN oscillators are given by equation
    (\ref{eq:FN general}). The parameters used in all simulations are
    $a=0.3$, $b=0.2$, $c=30$. (A) shows the trajectory of the
    ``membrane potential'' of a noise-free oscillator and (B) depicts
    the frequency spectrum of this trajectory computed by Fast Fourier
    Transformation. (C) and (D) present the trajectory (respectively
    the frequency spectrum) of a \emph{noisy uncoupled} oscillator
    ($\sigma=10$). (E) and (F) show the trajectory (respectively the
    frequency spectrum) of a \emph{noisy synchronized} oscillator
    within an all-to-all network ($\sigma=10$, $k_{ij}=5$,
    $n=200$). Note the temporal and frequential similarities between a
    noise-free oscillator and a noisy synchronized one. For instance,
    the main frequency and the first harmonics are very similar in the
    two frequency spectra. In contrast, the frequency spectrum of a
    noisy uncoupled oscillator shows no clear harmonics.}
  \label{fig:fn general}
\end{figure}

\begin{figure}[h!] 
  \centering
  \includegraphics[width=0.99\linewidth]{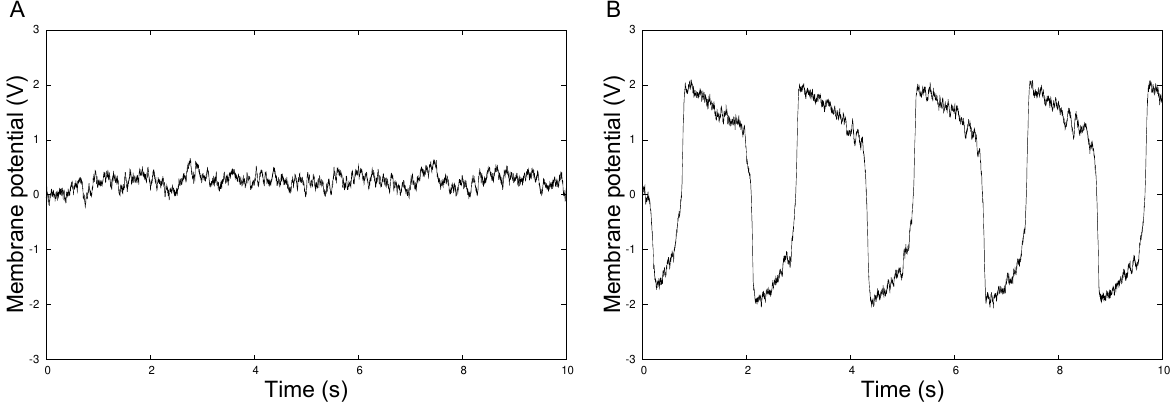}
  \caption{
    {\bf ``Spatial mean'' of FN oscillators.}
    Note that the same set of random initial conditions was used in
    the two plots.  (A) shows the average ``membrane potential''
    computed over $n=200$ \emph{noisy uncoupled} oscillators
    ($\sigma=10$). (B) shows the average ``membrane potential''
    computed over $n=200$ \emph{noisy synchronized} oscillators within
    an all-to-all network ($\sigma=10$, $k_{ij}=5$). Observe that, in
    the first plot, the average trajectory of uncoupled oscillators
    carries essentially no information, while in the second plot, the
    average trajectory of synchronized oscillators is very similar to
    a noise-free one.}
  \label{fig:fn mean}
\end{figure}

\begin{figure}[h!]
  \centering
  \includegraphics[width=0.99\linewidth]{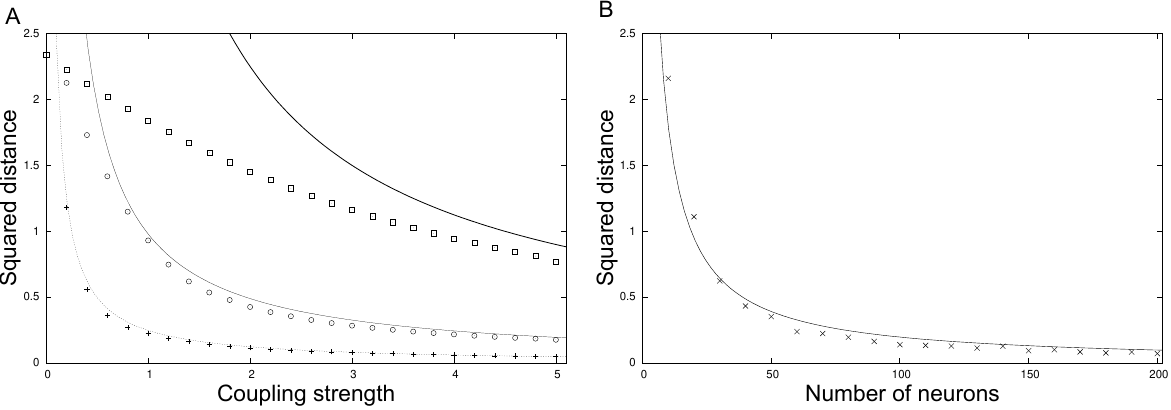}
  \caption{{\bf Asymptotic appraisal of the theoretical bounds.}
    Note that the experimental expectations were computed assuming the
    ergodic hypothesis. (A) Expectation of the average squared distance
    between the $v_i$'s and $\vmass$ (given by
    $\frac{1}{n}\expect\sum_i(v_i-\vmass)^2$) as a function of the
    coupling strength $k_{ij}$ ($\sigma=10$). Theoretical bound
    $\frac{(n-1)\sigma^2}{n^2k_{ij}}$ (cf equations~(\ref{eq:bound
      general}) and (\ref{eq:rho})) for $n=10$ (bold line), for $n=50$
    (plain line), for $n=200$ (dashed line); simulation results for
    $n=10$ (squares), for $n=50$ (triangles), for $n=200$ (crosses). (B)
    Expected squared distance between a noisy synchronized oscillator
    and its observer (given by $(\vobs-v_i)^2$) as a function of $n$
    ($\sigma=10$, $k_{ij}=5$). The bound
    $\frac{(n-1)\sigma^2}{n^2k_{ij}}$ was plotted in plain line and the
    simulation results were represented by crosses.
  }
  \label{fig:distance}
\end{figure}

\begin{figure}[h!]
  \centering
  \includegraphics[width=0.99\linewidth]{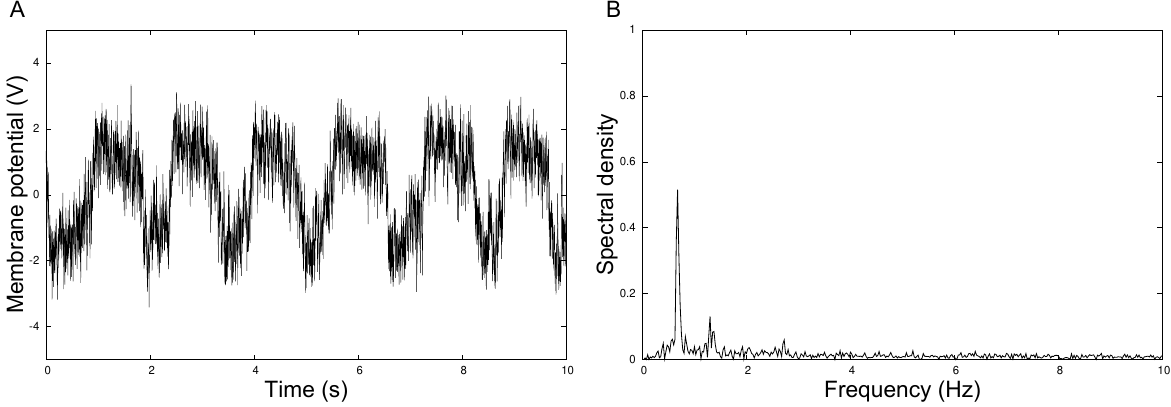}
  \caption{{\bf Simulation for a probabilistic symmetric network}
    ($n=200$, $p=0.1$, $\sigma=10$, $k_{ij}=5$). (A) shows the
    trajectory of the ``membrane potential'' of an oscillator in the
    network. (B) shows its frequency spectrum. Compare these two plots
    with those in Fig. 1.}
  \label{fig:proba}
\end{figure}

\begin{figure}[h!] 
  \centering
  \includegraphics[width=0.99\linewidth]{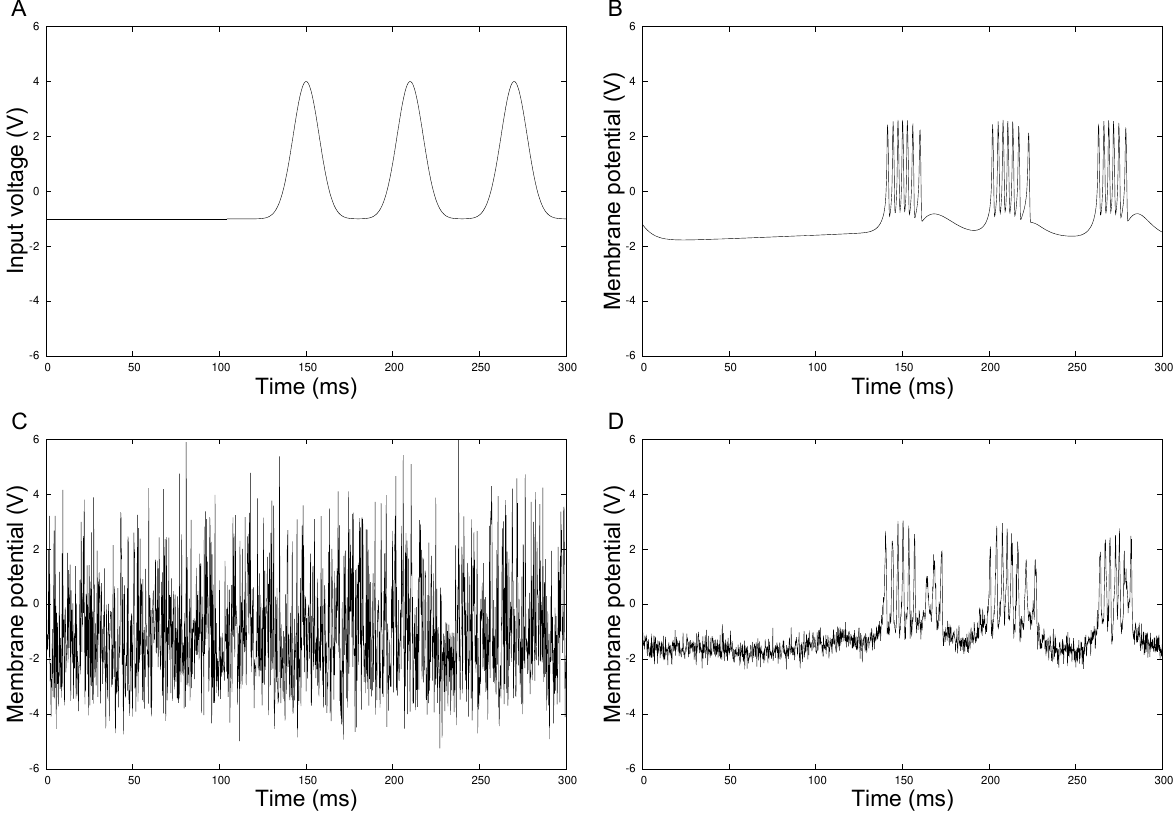}
  \caption{{\bf Simulation of Hindmarsh-Rose oscillators with time
      varying inputs.}
    (A) The time-varying input voltage. (B) Trajectory of the
    ``membrane potential'' of a noise-free oscillator. (C) Trajectory
    of a \emph{noisy uncoupled} oscillator. (D) Trajectory of a
    \emph{noisy synchronized} oscillator ($n=200$, $\sigma=10$,
    $k_{ij}=5$).}
\end{figure}



\end{document}